\documentclass[10pt,english,two column]{IEEEtran}
\usepackage[T1]{fontenc}
\usepackage[latin9]{inputenc}
\usepackage{float}
\usepackage{amsmath}
\usepackage{amsthm}
\usepackage{amssymb}
\usepackage{graphicx}
\usepackage{setspace}
\makeatletter

\floatstyle{ruled}
\newfloat{algorithm}{tbp}{loa}
\providecommand{\algorithmname}{Algorithm}
\floatname{algorithm}{\protect\algorithmname}

\theoremstyle{plain}

\theoremstyle{plain}

\ifCLASSINFOpdf
\else
\fi
\usepackage{babel}
\usepackage{algorithmic}

\makeatother

\usepackage{babel}
\providecommand{\propositionname}{Proposition}
\providecommand{\theoremname}{Theorem}

\begin{document}

\title{An Efficient Rate-Splitting Multiple Access Scheme for the Downlink
of C-RAN Systems}

\author{Daesung Yu, Junbeom Kim and Seok-Hwan Park \thanks{This work was supported by Basic Science Research Program through
the National Research Foundation of Korea (NRF) grants funded by the
Ministry of Education {[}NRF-2018R1D1A1B07040322, NRF-2019R1A6A1A09031717{]}.
(Corresponding author: Seok-Hwan Park.)

The authors are with the Division of Electronic Engineering, Chonbuk
National University, Jeonju, Korea (email: \{imcreative93, junbeom,
seokhwan\}@jbnu.ac.kr).}}
\maketitle
\begin{abstract}
This work studies the optimization of rate-splitting multiple access
(RSMA) transmission technique for a cloud radio access network (C-RAN)
downlink system. Main idea of RSMA is to split the message for each
user equipment (UE) to private and common messages and perform superposition
coding at transmitters so as to enable flexible decoding at receivers.
It is challenging to implement ideal RSMA scheme particularly when
there are many UEs, since the number of common signals exponentially
increases with the number of UEs. An efficient RSMA scheme is hence
proposed that uses a linearly increasing number of common signals
whose decoding UEs are selected using hierarchical clustering. Via
numerical results, we show the performance gains of the proposed RSMA
scheme over conventional space-division multiple access (SDMA) and
non-orthogonal multiple access (NOMA) schemes as well as over a conventional
RSMA scheme that uses a single common signal.\end{abstract}


\theoremstyle{theorem}
\newtheorem{theorem}{Theorem}
\theoremstyle{proposition}
\newtheorem{proposition}{Proposition}
\theoremstyle{lemma}
\newtheorem{lemma}{Lemma}
\theoremstyle{corollary}
\newtheorem{corollary}{Corollary}
\theoremstyle{definition}
\newtheorem{definition}{Definition}
\theoremstyle{remark}
\newtheorem{remark}{Remark}

\section{Introduction\label{sec:Introduction}}

Cloud radio access network (C-RAN) is promised to meet the requirement
of massive connectivity by means of centralized signal processing
at baseband processing unit (BBU) pool \cite{Simeone-et-al:JCN}.
In \cite{Park-et-al:TSP13}, the signal processing optimization for
C-RAN downlink systems was tackled under the assumption that user
equipments (UEs) perform single-user decoding. As in \cite{Mao-et-al:Eurasip},
we refer to this approach as space-division multiple access (SDMA).
To further improve the connectivity support level in C-RAN systems,
an application of non-orthogonal multiple access (NOMA) technique
was discussed in \cite{Lee-et-al:WCL18}, where each UE performs successive
interference cancellation (SIC) decoding until its target message
is obtained. However, the NOMA scheme does not always guarantee better
performance than SDMA, since the achievable rates need to be limited
for successful SIC decoding.

In another efficient multiple access technique, referred to as \textit{rate-splitting
multiple access} (RSMA) \cite{Mao-et-al:Eurasip}, the message intended
for each UE is split into a private message and possibly multiple
common messages which are simultaneously transmitted on the downlink
by means of superposition coding. Each UE decodes and cancels a predetermined
set of common messages and then decodes its private message. The advantages
of RSMA in terms of rate region or sum-rate were discussed in \cite{Mao-et-al:Eurasip}
and \cite{Mao-et-al:SPAWC18} for a multi-user downlink system. The
discussion was extended in \cite{Mao-et-al:arXiv18} to a coordinated
multi-cell system in which BSs can perfectly cooperate, and in \cite{Ahmad}
to a C-RAN system under the assumption that, as in \cite{Dahrouj},
each UE has a single common message, and different common messages
are independently encoded.

In this work, motivated by the RSMA schemes in \cite{Mao-et-al:Eurasip}
and \cite{Mao-et-al:arXiv18} for multi-user systems, we propose an
efficient RSMA transmission scheme for a C-RAN downlink system. We
first apply the RSMA schemes \cite{Mao-et-al:Eurasip}\cite{Mao-et-al:arXiv18}
to a C-RAN system, whereby unlike the RSMA scheme in \cite{Ahmad},
each UE may have multiple common messages depending on the decoding
UEs, and the common messages decoded by the same UEs are jointly encoded.
We tackle the joint optimization of rate-splitting, precoding and
fronthaul compression strategies using the weighted minimum mean squared
error (WMMSE) approach \cite{Zhou-Yu}. It is challenging to implement
the ideal RSMA scheme \cite{Mao-et-al:Eurasip} due to the number
of common signals, and hence of precoding vectors to be optimized,
exponentially increasing with the number of UEs. Therefore, we propose
an efficient RSMA scheme that uses a linearly increasing number of
common signals whose decoding UEs are selected using hierarchical
clustering \cite{Rokach}\cite{Wiki}. Via numerical results, we show
the performance gains of the proposed scheme compared to conventional
SDMA and NOMA schemes as well as to a conventional RSMA scheme \cite{Mao-et-al:SPAWC18}
that uses a single common signal.

\section{System Model\label{sec:System-Model}}

We consider a C-RAN downlink system with a BBU, $N_{R}$ remote radio
heads (RRHs) and $N_{U}$ single-antenna UEs. The BBU is connected
to RRH $i$, which uses $n_{R,i}$ antennas, through a fronthaul link
of capacity $C_{i}$ bit/symbol. We define the sets $\mathcal{N}_{R}\triangleq\{1,2,\ldots,N_{R}\}$
and $\mathcal{N}_{U}\triangleq\{1,2,\ldots,N_{U}\}$ of the RRHs'
and UEs' indices, respectively, and the total number $n_{R}\triangleq\sum_{i\in\mathcal{N}_{R}}n_{R,i}$
of RRHs' antennas.

On the downlink, the signal $y_{k}$ received by UE $k$ is given
by $y_{k}=\sum\nolimits _{i\in\mathcal{N}_{R}}\mathbf{h}_{k,i}^{H}\mathbf{x}_{i}+z_{k}$,
where $\mathbf{h}_{k,i}\in\mathbb{C}^{n_{R,i}\times1}$ denotes the
channel vector between RRH $i$ and UE $k$; $\mathbf{x}_{i}\in\mathbb{C}^{n_{R,i}\times1}$
represents the transmit signal vector of RRH $i$; and $z_{k}\sim\mathcal{CN}(0,\sigma_{k}^{2})$
is the additive noise at UE $k$. For convenience, we also define
the stacked channel vector $\mathbf{h}_{k}\triangleq[\mathbf{h}_{k,1}^{H}\,\ldots\,\mathbf{h}_{k,N_{R}}^{H}]^{H}$
from all the RRHs to UE $k$. For each RRH $i$, a transmit power
constraint $\mathtt{E}\left\Vert \mathbf{x}_{i}\right\Vert ^{2}\leq P_{i}$
is imposed. We assume that the coding blocklength is sufficiently
large and the channels remain constant within each block.

\section{Optimizing RSMA for C-RAN Downlink}

In this section, we discuss the design of the RSMA scheme, which was
studied in \cite{Mao-et-al:Eurasip}\cite{Mao-et-al:SPAWC18}\cite{Mao-et-al:arXiv18}
for multi-user or coordinated multi-cell systems, for the downlink
of C-RAN system. To elaborate, we first define sets $\mathcal{S}_{1},\mathcal{S}_{2},\ldots,\mathcal{S}_{L}$
used for grouping common signals which satisfy the following conditions:
\textit{i)} $\mathcal{S}_{l}\subseteq\mathcal{N}_{U}$ for all $l\in\mathcal{L}\triangleq\{1,2,\ldots,L\}$;
\textit{ii)} $\mathcal{S}_{l}\neq\mathcal{S}_{m}$ for all $l\neq m\in\mathcal{L}$;
and \textit{iii)} $|\mathcal{S}_{l}|\geq2$ for all $l\in\mathcal{L}$.
We assume that the BBU splits the message $M_{k}$ of rate $R_{k}$
for each UE $k$ into a private message $M_{p,k}$ and common messages
$\{M_{c,k,l}\}_{l\in\mathcal{L}_{k}}$, where $\mathcal{L}_{k}=\{l\in\mathcal{L}|k\in\mathcal{S}_{l}\}$
collects the indices of the sets that contain UE $k$. Accordingly,
the private message $M_{p,k}$ of rate $R_{p,k}$ is decoded only
by UE $k$, while the common message $M_{c,k,l}$ of rate $R_{c,k,l}$
is decoded by the UEs in $\mathcal{S}_{l}$. In what follows, we address
the optimization of rate-splitting, precoding and fronthaul compression
strategies for fixed sets $\mathcal{S}_{1},\mathcal{S}_{2},\ldots,\mathcal{S}_{L}$,
and then conventional and proposed designs of $\mathcal{S}_{1},\mathcal{S}_{2},\ldots,\mathcal{S}_{L}$
are discussed.

\subsection{Optimizing RSMA for fixed $\mathcal{S}_{1},\mathcal{S}_{2},\ldots,\mathcal{S}_{L}$}

The BBU encodes each private message $M_{p,k}$ to a baseband signal
$s_{p,k}\sim\mathcal{CN}(0,1)$, and the common messages $\{M_{c,k,l}\}_{k\in\mathcal{S}_{l}}$,
which are decoded by the same UEs $S_{l}$, are jointly encoded to
a signal $s_{c,l}\sim\mathcal{CN}(0,1)$\footnote{Encoding distinct common messages to independent codewords was studied
in \cite{Ahmad}, but in this work, we focus on joint encoding.}. The encoded signals are then linearly precoded as $\tilde{\mathbf{x}}=\sum\nolimits _{k\in\mathcal{N}_{U}}\mathbf{v}_{p,k}s_{p,k}+\sum\nolimits _{l\in\mathcal{L}}\mathbf{v}_{c,l}s_{c,l}$,
where $\mathbf{v}_{p,k}\in\mathbb{C}^{n_{R}\times1}$ and $\mathbf{v}_{c,l}\in\mathbb{C}^{n_{R}\times1}$
denote the precoding vectors for the signals $s_{p,k}$ and $s_{c,l}$,
respectively; and $\tilde{\mathbf{x}}=[\tilde{\mathbf{x}}_{1}^{H}\,\tilde{\mathbf{x}}_{2}^{H}\,\cdots\,\tilde{\mathbf{x}}_{N_{R}}^{H}]\in\mathbb{C}^{n_{R}\times1}$
represents the precoding output signal for all the RRHs with the $i$th
subvector $\tilde{\mathbf{x}}_{i}\in\mathbb{C}^{n_{R,i}\times1}$
to be transferred to RRH $i$. Note that the vector $\tilde{\mathbf{x}}_{i}$
can be represented as $\tilde{\mathbf{x}}_{i}=\mathbf{E}_{i}^{H}\tilde{\mathbf{x}}$,
where a matrix $\mathbf{E}_{i}\in\mathbb{C}^{n_{R}\times n_{R,i}}$
is filled with zeros except for the rows from $\sum_{j=1}^{i-1}n_{R,j}+1$
to $\sum_{j=1}^{i}n_{R,j}$ being an identity matrix.

The BBU quantizes and compresses the precoding output vector $\tilde{\mathbf{x}}_{i}$
obtaining a quantized version $\mathbf{x}_{i}$, which is transferred
to RRH $i$ on the fronthaul link and transmitted on the downlink
channel. Following related works \cite{Park-et-al:TSP13}\cite{Lee-et-al:WCL18}\cite{Zhou-Yu},
we model the impact of quantization as $\mathbf{x}_{i}=\tilde{\mathbf{x}}_{i}+\mathbf{q}_{i}$
with the quantization noise $\mathbf{q}_{i}\sim\mathcal{CN}(\mathbf{0},\mathbf{\Omega}_{i})$.
The covariance matrix $\mathbf{\Omega}_{i}$ should satisfy the condition:
$g_{i}(\mathbf{v},\mathbf{\Omega})=I(\tilde{\mathbf{x}}_{i};\mathbf{x}_{i})=\Phi(\mathbf{E}_{i}^{H}(\sum\nolimits _{k\in\mathcal{N}_{U}}\!\mathbf{v}_{p,k}\mathbf{v}_{p,k}^{H}\!+\!\sum\nolimits _{l\in\mathcal{L}}\!\mathbf{v}_{c,\mathcal{S}}\mathbf{v}_{c,\mathcal{S}}^{H})\mathbf{E}_{i},\mathbf{\Omega}_{i})\leq C_{i}$,
where we have defined $\Phi(\mathbf{A},\mathbf{B})=\log_{2}\det(\mathbf{A}+\mathbf{B})-\log_{2}\det(\mathbf{B})$,
$\mathbf{v}\triangleq\{\mathbf{v}_{p,k}\}_{k\in\mathcal{N}_{U}}\cup\{\mathbf{v}_{c,l}\}_{l\in\mathcal{L}}$
and $\mathbf{\Omega}\triangleq\{\mathbf{\Omega}_{i}\}_{i\in\mathcal{N}_{R}}$.

To describe the SIC decoding process at UE $k$, we define a permutation
$\pi_{k}:\{1,\ldots,L_{k}\}\rightarrow\mathcal{L}_{k}$ that represents
the decoding order of $L_{k}=|\mathcal{L}_{k}|$ common signals at
UE $k$. Accordingly, UE $k$ decodes the common signals $\{s_{c,l}\}_{l\in\mathcal{L}_{k}}$
and private signal $s_{p,k}$ by SIC decoding with the order $s_{c,\pi_{k}(1)}\rightarrow s_{c,\pi_{k}(2)}\rightarrow\ldots\rightarrow s_{c,\pi_{k}(L_{k})}\rightarrow s_{p,k}$.
In this work, we randomly fix $\pi_{k}$ while satisfying $|\mathcal{S}_{\pi_{k}(m)}|\geq|\mathcal{S}_{\pi_{k}(n)}|$
for all $m<n\in\{1,\ldots,L_{k}\}$ without claim of optimality. Under
this assumption, the constraints on the achievable rates of the private
and common messages can be stated as
\begin{align}
 & R_{p,k}\leq f_{p,k}\left(\mathbf{v},\mathbf{\Omega}\right),\label{eq:achievable-rate-private}\\
\mathrm{and}\,\, & \sum\nolimits _{k\in\mathcal{S}_{l}}R_{c,k,l}\leq\min\nolimits _{k\in\mathcal{S}_{l}}\,f_{c,l,k}\left(\mathbf{v},\mathbf{\Omega}\right),
\end{align}
where the functions $f_{p,k}(\mathbf{v},\mathbf{\Omega})$ and $f_{c,l,k}(\mathbf{v},\mathbf{\Omega})$
are defined as $\!\!f_{p,k}\left(\mathbf{v},\mathbf{\Omega}\right)\!=\!I\!(s_{p,k};y_{k}|\{s_{c,l}\}_{l\in\mathcal{L}_{k}})\!=\!\Phi\!(|\mathbf{h}_{k}^{H}\mathbf{v}_{p,k}|^{2},\!\nu_{p,k}\left(\mathbf{v},\mathbf{\Omega}\right))$
and $\!\!f_{c,l,k}\left(\mathbf{v},\mathbf{\Omega}\right)\!=\!I\!(s_{c,l};y_{k}|\{s_{c,\pi(m)}\}_{m=\pi_{k}^{-1}(l)+1}^{L_{k}})\!=\!\Phi\!(|\mathbf{h}_{k}^{H}\mathbf{v}_{c,l}|^{2},\nu_{c,l,k}\left(\mathbf{v},\mathbf{\Omega}\right))$
with $\bar{\mathbf{\Omega}}\triangleq\mathrm{diag}(\{\mathbf{\Omega}_{i}\}_{i\in\mathcal{N}_{R}})$,
$\nu_{p,k}(\mathbf{v},\mathbf{\Omega})\triangleq\sum_{l\in\mathcal{L}\setminus\mathcal{L}_{k}}\!|\mathbf{h}_{k}^{H}\mathbf{v}_{c,l}|^{2}+\sum_{m\in\mathcal{N}_{U}\setminus\{k\}}\!|\mathbf{h}_{k}^{H}\mathbf{v}_{p,m}|^{2}+\mathbf{h}_{k}^{H}\bar{\mathbf{\Omega}}\mathbf{h}_{k}+\sigma_{k}^{2}$
and $\nu_{c,l,k}(\mathbf{v},\mathbf{\Omega})\triangleq\sum_{m=\pi_{l}^{-1}(k)+1}^{L_{k}}\!|\mathbf{h}_{k}^{H}\mathbf{v}_{c,\pi(m)}|^{2}+\sum_{m\in\mathcal{L}\setminus\mathcal{L}_{k}}\!|\mathbf{h}_{k}^{H}\mathbf{v}_{c,m}|^{2}+\sum_{m\in\mathcal{N}_{U}}\!|\mathbf{h}_{k}^{H}\mathbf{v}_{p,m}|^{2}+\mathbf{h}_{k}^{H}\bar{\mathbf{\Omega}}\mathbf{h}_{k}+\sigma_{k}^{2}$.

We now tackle the problem of jointly optimizing the rate-splitting
variables $\mathbf{R}=\{R_{p,k}\}_{k\in\mathcal{N}_{U}}\cup\{R_{c,k,l}\}_{k\in\mathcal{N}_{U},l\in\mathcal{L}_{k}}$,
the precoding vectors $\mathbf{v}$ and the quantization noise covariance
matrices $\mathbf{\Omega}$ to maximize the minimum-UE rate $R_{\min}\triangleq\min_{k\in\mathcal{N}_{U}}R_{k}$.
Here, we focus on the minimum-UE rate to guarantee the fairness among
the UEs \cite{Timotheou}\footnote{The $\alpha$-fairness metric is also widely used to guarantee fairness
\cite{Xu}, but we focus on the minimum-UE rate in this work.}. We can mathematically formulate the problem as\begin{subequations}\label{eq:problem}
\begin{align}
\underset{\mathbf{v},\mathbf{\Omega},\mathbf{R}}{\mathrm{maximize}}\, & \min_{k\in\mathcal{N}_{U}}\left\{ R_{p,k}+\sum\nolimits _{l\in\mathcal{L}_{k}}R_{c,k,l}\right\} \label{eq:problem-objective}\\
\mathrm{s.t.}\,\, & R_{p,k}\leq f_{p,k}\left(\mathbf{v},\mathbf{\Omega}\right),\,\,k\in\mathcal{N}_{U},\label{eq:problem-rate-linear}\\
 & \sum_{k\in\mathcal{S}_{l}}R_{c,k,l}\leq f_{c,l,k}\left(\mathbf{v},\mathbf{\Omega}\right),\,\,l\in\mathcal{L},\,k\in\mathcal{S}_{l},\label{eq:problem-rate-NOMA}\\
 & g_{i}\left(\mathbf{v},\mathbf{\Omega}\right)\leq C_{i},\,\,i\in\mathcal{N}_{R},\label{eq:problem-fronthaul-constraint}\\
 & p_{i}\left(\mathbf{v},\mathbf{\Omega}\right)\leq P_{i},\,i\in\mathcal{N}_{R},\label{eq:problem-power}
\end{align}
\end{subequations}where we have defined $p_{i}(\mathbf{v},\mathbf{\Omega})=\mathrm{tr}(\mathrm{cov}_{\mathbf{x}_{i}}\left(\mathbf{v},\mathbf{\Omega}\right))$
with $\mathrm{cov}_{\mathbf{x}_{i}}\left(\mathbf{v},\mathbf{\Omega}\right)=\mathtt{E}[\mathbf{x}_{i}\mathbf{x}_{i}^{H}]=\mathbf{E}_{i}^{H}(\sum\nolimits _{k\in\mathcal{N}_{U}}\mathbf{v}_{p,k}\mathbf{v}_{p,k}^{H}+\sum\nolimits _{l\in\mathcal{L}}\mathbf{v}_{c,l}\mathbf{v}_{c,l}^{H})\mathbf{E}_{i}+\mathbf{\Omega}_{i}$.

Since it is difficult to solve the problem (\ref{eq:problem}), we
adopt the WMMSE based approach \cite{Zhou-Yu}. We first note that
the rate functions $f_{p,k}(\mathbf{v},\mathbf{\Omega})$ and $f_{c,l,k}(\mathbf{v},\mathbf{\Omega})$
are lower bounded as
\begin{align}
f_{p,k}\left(\mathbf{v},\mathbf{\Omega}\right) & \geq\widetilde{f}_{p,k}\left(\mathbf{v},\mathbf{\Omega},u_{p,k},w_{p,k}\right)\label{eq:LB-rate-private}\\
 & \triangleq\log_{2}w_{p,k}+\frac{1}{\ln2}\left(1-w_{p,k}e_{p,k}\left(\mathbf{v},\mathbf{\Omega},u_{p,k}\right)\right),\nonumber \\
f_{c,l,k}\left(\mathbf{v},\mathbf{\Omega}\right) & \geq\widetilde{f}_{c,l,k}\left(\mathbf{v},\mathbf{\Omega},u_{c,l,k},w_{c,l,k}\right)\label{eq:LB-rate-common}\\
 & \!\!\!\!\triangleq\log_{2}w_{c,l,k}+\!\frac{1}{\ln\!2}\left(1\!-w_{c,l,k}e_{c,l,k}\left(\mathbf{v},\mathbf{\Omega},u_{c,l,k}\right)\right),\nonumber
\end{align}
for arbitrary receive filters $u_{p,k},u_{c,l,k}$ and nonnegative
weights $w_{p,k},w_{c,l,k}\geq0$. Here, the error variance functions
are defined as $e_{p,k}(\mathbf{v},\mathbf{\Omega},u_{p,k})\triangleq\mathtt{E}[|u_{p,k}^{H}y_{k}-s_{p,k}|^{2}]=|u_{p,k}^{H}\mathbf{h}_{k}^{H}\mathbf{v}_{p,k}-1|{}^{2}+|u_{p,k}|^{2}\nu_{p,k}(\mathbf{v},\mathbf{\Omega})$
and $e_{c,l,k}(\mathbf{v},\mathbf{\Omega},u_{c,l,k})\triangleq\mathtt{E}[|u_{c,l,k}^{H}y_{k}-s_{c,l}|^{2}]=|u_{c,l,k}^{H}\mathbf{h}_{k}^{H}\mathbf{v}_{c,l}-1|{}^{2}+|u_{c,l,k}|^{2}\nu_{c,l,k}\left(\mathbf{v},\mathbf{\Omega}\right)$.
It was shown in \cite{Zhou-Yu} that the lower bounds (\ref{eq:LB-rate-private})
and (\ref{eq:LB-rate-common}) are tight when the receive filters
$u_{p,k},u_{c,l,k}$ and weights $w_{p,k},w_{c,l,k}$ become
\begin{align}
u_{p,k} & =\frac{\mathbf{h}_{k}^{H}\mathbf{v}_{p,k}}{\nu_{p,k}\left(\mathbf{v},\mathbf{\Omega}\right)},\,\,u_{c,l,k}=\frac{\mathbf{h}_{k}^{H}\mathbf{v}_{c,l}}{\nu_{c,l,k}\left(\mathbf{v},\mathbf{\Omega}\right)},\label{eq:optimal-rx-filters}\\
w_{p,k} & =\frac{1}{e_{p,k}\left(\mathbf{v},\mathbf{\Omega},u_{p,k}\right)},\,\,w_{c,l,k}=\frac{1}{e_{c,l,k}\left(\mathbf{v},\mathbf{\Omega},u_{c,l,k}\right)}.\label{eq:optimal-weights}
\end{align}

To handle the non-convex constraint (\ref{eq:problem-fronthaul-constraint}),
we consider an upper bound on the left-hand side (LHS) as
\begin{align}
 & g_{i}\left(\mathbf{v},\mathbf{\Omega},\mathbf{\Sigma}_{i}\right)\leq\widetilde{g}_{i}\left(\mathbf{v},\mathbf{\Omega},\mathbf{\Sigma}_{i}\right)\triangleq\log_{2}\!\det(\mathbf{\Sigma}_{i})\!\label{eq:fronthaul-constraint-alternative}\\
 & +\frac{1}{\ln2}\!\left(\mathrm{tr}\!\left(\mathbf{\Sigma}_{i}^{-1}\mathrm{cov}_{\mathbf{x}_{i}}\!\left(\mathbf{v},\!\mathbf{\Omega}\right)\right)\!-\!n_{R,i}\right)\!-\!\log_{2}\det(\mathbf{\Omega}_{i}),\nonumber
\end{align}
which holds for any positive definite matrix $\mathbf{\Sigma}_{i}$.
The optimal matrix $\mathbf{\Sigma}_{i}$ which makes the lower bound
(\ref{eq:fronthaul-constraint-alternative}) tight is
\begin{align}
 & \mathbf{\Sigma}_{i}=\mathrm{cov}_{\mathbf{x}_{i}}\left(\mathbf{v},\mathbf{\Omega}\right).\label{eq:optimal-Sigma}
\end{align}

Based on the above discussion, we tackle the problem (\ref{eq:problem})
by replacing the functions $f_{p,k}(\mathbf{v},\mathbf{\Omega})$,
$f_{c,l,k}(\mathbf{v},\mathbf{\Omega})$ and $g_{i}(\mathbf{v},\mathbf{\Omega})$
with $\widetilde{f}_{p,k}(\mathbf{v},\mathbf{\Omega},u_{p,k},w_{p,k})$,
$\widetilde{f}_{c,l,k}(\mathbf{v},\mathbf{\Omega},u_{c,l,k},w_{c,l,k})$
and $\widetilde{g}_{i}(\mathbf{v},\mathbf{\Omega},\mathbf{\Sigma}_{i})$,
respectively, and adding $\mathbf{u}\triangleq\{u_{p,k}\}_{k\in\mathcal{N}_{U}}\cup\{u_{c,l,k}\}_{l\in\mathcal{L},k\in\mathcal{S}_{l}}$,
$\mathbf{w}\triangleq\{w_{p,k}\}_{k\in\mathcal{N}_{U}}\cup\{w_{c,l,k}\}_{l\in\mathcal{L},k\in\mathcal{S}_{l}}$
and $\mathbf{\Sigma}\triangleq\{\mathbf{\Sigma}_{i}\}_{i\in\mathcal{N}_{R}}$
to the set of optimization variables. It is straightforward to see
that, in the so-obtained problem which we refer to as \textit{WMMSE
problem}, optimizing the variables $\left\{ \mathbf{R},\mathbf{v},\mathbf{\Omega}\right\} $
for fixed $\{\mathbf{u},\mathbf{w},\mathbf{\Sigma}\}$ is a convex
problem. Also, the optimal values of $\{\mathbf{u},\mathbf{w},\mathbf{\Sigma}\}$
for fixed $\left\{ \mathbf{v},\mathbf{\Omega}\right\} $ are given
as (\ref{eq:optimal-rx-filters}), (\ref{eq:optimal-weights}) and
(\ref{eq:optimal-Sigma}). Therefore, we can derive an iterative algorithm
which gives monotonically non-decreasing minimum-UE rates with respect
to the number of iterations. The detailed algorithm is described in
Algorithm 1.

\begin{algorithm}
\caption{WMMSE based algorithm for problem (\ref{eq:problem})}

\textbf{\footnotesize{}1}\textbf{ Initialize:} $t\leftarrow1$, $\mathbf{v}^{[t]}$,
$\mathbf{\Omega}^{[t]}$;

\textbf{\footnotesize{}2}\textbf{ repeat}

\textbf{\footnotesize{}3}~~~~$t\leftarrow t+1$;

\textbf{\footnotesize{}4}~~~~update $(\mathbf{u}^{[t]},\mathbf{w}^{[t]},\mathbf{\Sigma}^{[t]})$
using (\ref{eq:optimal-rx-filters}), (\ref{eq:optimal-weights})
and (\ref{eq:optimal-Sigma}) for fixed

~~~~~$\mathbf{v}\leftarrow\mathbf{v}^{[t-1]}$ and $\mathbf{\Omega}\leftarrow\mathbf{\Omega}^{[t-1]}$;

\textbf{\footnotesize{}5}~~~~update $(\mathbf{v}^{[t]},\mathbf{\Omega}^{[t]})$
by solving the WMMSE problem for

~~~~~fixed $\mathbf{u}\leftarrow\mathbf{u}^{[t]}$, $\mathbf{w}\leftarrow\mathbf{w}^{[t]}$
and $\mathbf{\Sigma}\leftarrow\mathbf{\Sigma}^{[t]}$;

\textbf{\footnotesize{}6} \textbf{until} $|R_{\min}^{[t]}-R_{\min}^{[t-1]}|\leq\epsilon$;
\end{algorithm}

The complexity of Algorithm 1 is given as the product of the number
of iterations and the complexity of each iteration. We observed from
simulation that Algorithm 1 converges within a few tens of iterations.
The complexity of each iteration is dominated by Step 5 which solves
a convex problem for updating $\mathbf{v}$ and $\mathbf{\Omega}$.
The complexity of solving a convex problem is known polynomial in
the problem size \cite[Ch. 11]{Boyd} which is here $\mathcal{O}(N_{R}(N_{U}+L))$
assuming that the numbers of RRH antennas are fixed. This analysis
suggests that the problem size has a significant impact on the complexity
of Algorithm 1 and it is important to develop an efficient RSMA scheme
that uses a reasonable number $L$ of common signals.

\subsection{Designing Sets $\mathcal{S}_{1},\mathcal{S}_{2},\ldots,\mathcal{S}_{L}$
for Common Signals\label{sub:Selecting-Sets-common-signals}}

We first review the conventional designs in \cite{Mao-et-al:Eurasip}
and \cite{Mao-et-al:SPAWC18}, and propose an efficient scheme based
on the hierarchical agglomerative clustering.

\subsubsection{Conventional Designs of $\mathcal{S}_{1},\mathcal{S}_{2},\ldots,\mathcal{S}_{L}$}

It was proposed in \cite{Mao-et-al:Eurasip} to use all possible subsets
for common signals, and each UE decodes the target common signals
in the descending order of the cardinality of the subsets. With this,
the number of common signals equals $L=2^{N_{U}}-1-N_{U}$, which
leads to the problem size of $\mathcal{O}(N_{R}2^{N_{U}})$ in Step
5 and makes it impractical to run Algorithm 1.

Instead, the work \cite{Mao-et-al:SPAWC18} proposed to use a single
common signal which is decoded by all UEs, i.e., $L=1$ with $\mathcal{S}_{1}=\mathcal{N}_{U}$.
We refer to the RSMA scheme optimized with this choice as \textit{RSMA
with a single common signal} (RSMA-SC). The problem size in Step 5
with RSMA-SC increases in $\mathcal{O}(N_{R}N_{U})$, which gives
significantly lower complexity than the scheme in \cite{Mao-et-al:Eurasip}.
In Sec. \ref{sec:Numerical-Results}, we will show that this simple
approach shows significant gains over the SDMA and NOMA schemes.

\subsubsection{Proposed Design of $\mathcal{S}_{1},\mathcal{S}_{2},\ldots,\mathcal{S}_{L}$}

We now propose an RSMA scheme that shows a notable gain compared to
the RSMA-SC scheme. Unlike RSMA-SC which uses a single common signal,
we propose to use $L=N_{U}-1$ common signals, since the problem size
of Step 5 with $L=N_{U}-1$ is still $\mathcal{O}(N_{R}N_{U})$. For
the choice of the sets $\mathcal{S}_{1},\mathcal{S}_{2},\ldots,\mathcal{S}_{L}$,
we consider the following insight: For given $\mathcal{S}_{l}$, the
precoding vector $\mathbf{v}_{c,l}$ is applied for a multicast transmission
towards the UEs in $\mathcal{S}_{l}$. Therefore, it would be desirable
to make the UEs in the same set $\mathcal{S}_{l}$ have similar channel
directions. To reflect this observation, we define a measure of dissimilarity
between a pair of UEs $k$ and $m$, $k\neq m$, as $d_{U,k,m}=1-|\mathbf{h}_{k}^{H}\mathbf{h}_{m}|/(||\mathbf{h}_{k}||\,||\mathbf{h}_{m}||)\in[0,1]$.
Note that the defined distance metric $d_{U,k,m}$ increases as the
channel vectors of UEs $k$ and $m$ are less aligned.

Based on the distance metric $d_{U,k,m}$, we perform the \textit{hierarchical
agglomerative clustering} \cite{Rokach}\cite{Wiki} of UEs, which
builds a dendrogram in a ``bottom-up'' approach. We begin with the
bottom layer which has $N_{U}$ clusters each being the corresponding
UE. Then, we successively move up the hierarchy with merging the pair
having the minimum inter-cluster distance until a single cluster remains.
When measuring the inter-cluster distance $d_{C}(\mathcal{A},\mathcal{B})$
between clusters $\mathcal{A}$ and $\mathcal{B}$ with $\mathcal{A},\mathcal{B}\subseteq\mathcal{N}_{U}$
and $\mathcal{A}\cap\mathcal{B}=\emptyset$, we adopt the complete-linkage
criterion, i.e., $d_{C}(\mathcal{A},\mathcal{B})=\max_{k\in\mathcal{A},m\in\mathcal{B}}d_{U,k,m}$.
Once a dendrogram is built which has $N_{U}-1$ layers, we obtain
the sets $\mathcal{S}_{1},\mathcal{S}_{2},\ldots,\mathcal{S}_{L}$
by cutting every layer, i.e., layers 1 to $N_{U}-1$ while excluding
single-cardinality clusters. To guarantee that the proposed scheme
performs at least as well as the RSMA-SC scheme, we also include the
whole UE set $\mathcal{N}_{U}$. Since cutting each layer produces
two subsets, we always have $L=2(N_{U}-1)+1-N_{U}=N_{U}-1$ common
signals in total, where the term $1-N_{U}$ comes to reflect that
we include the whole UE set $\mathcal{N}_{U}$ and exclude single-cardinality
sets $\{k\}$, $k\in\mathcal{N}_{U}$. The complexity of the hierarchical
clustering algorithms is known to be $\mathcal{O}(N^{3})$ \cite{Wiki},
where $N$ is the number of observations which here equals $N=N_{U}$.

\section{Numerical Results\label{sec:Numerical-Results}}

For simulation, we assume that the RRHs and UEs are uniformly located
within a circular area of radius 100 m, and as in \cite{Dai-Yu:Access14},
we adopt the path-loss model $128.1+37.6\log_{10}d$, where $d$ denotes
the distance between the two nodes in km. We also consider small-scale
Rayleigh fading and log-normal shadowing with standard deviation equal
to 8 dB. The channel bandwidth and the noise power spectral density
are set to 10 MHz and -169 dBm/Hz, respectively. For comparison, we
present the performance of the conventional SDMA \cite{Park-et-al:TSP13}
and NOMA schemes \cite{Lee-et-al:WCL18} as well as of the RSMA-SC
scheme \cite{Mao-et-al:SPAWC18} with $L=1$ and $\mathcal{S}_{1}=\mathcal{N}_{U}$.
To validate the effectiveness of the proposed design of sets $\mathcal{S}_{1},\mathcal{S}_{2},\ldots,\mathcal{S}_{L}$
based on the hierarchical clustering, we also show the performance
of the RSMA scheme with $L=N_{U}-1$ common signals, where the $l$th
set $\mathcal{S}_{l}$ randomly picks $l+1$ UEs. We refer to this
scheme as \textit{RSMA with random common signals} (RSMA-RC).

\begin{figure}
\centering\includegraphics[width=8cm,height=6.5cm]{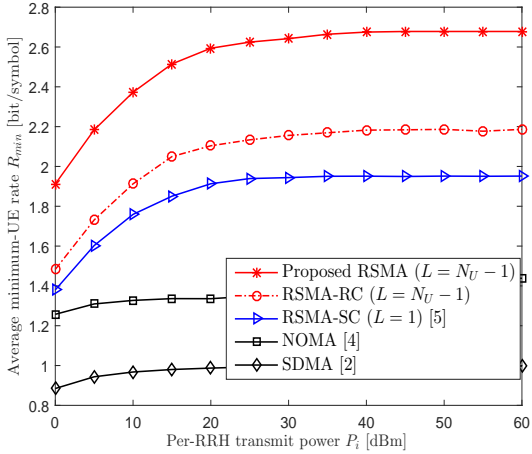}\caption{{\footnotesize{}\label{fig:graph-asSNR}Average minimum-UE rate $R_{\min}$
versus per-RRH transmit power $P_{i}$ ($N_{R}=4$, $N_{U}=8$, $n_{R,i}=1$
and $C_{i}=10$).}}
\end{figure}

In Fig. \ref{fig:graph-asSNR}, we plot the average minimum-UE rate
versus the transmit power $P_{i}$ for $N_{R}=4$, $N_{U}=8$, $n_{R,i}=1$
and $C_{i}=10$. We first note that the performance gap among different
schemes is more pronounced at larger transmit powers $P_{i}$ due
to the reduced impact of noise signals. The figure also shows that
the conventional RSMA-SC scheme \cite{Mao-et-al:SPAWC18} shows significant
gains over the NOMA and SDMA schemes using only a single common signal.
The RSMA-RC scheme shows a further gain by using $L=N_{U}-1$ common
signals, but the gain compared to RSMA-SC is minor, since the sets
$\mathcal{S}_{1},\mathcal{S}_{2},\ldots,\mathcal{S}_{L}$ are randomly
chosen. However, if we use the same number of common signals with
carefully chosen sets for the proposed RSMA, the performance gain
becomes considerable.

\begin{figure}
\centering\includegraphics[width=8cm,height=6.5cm]{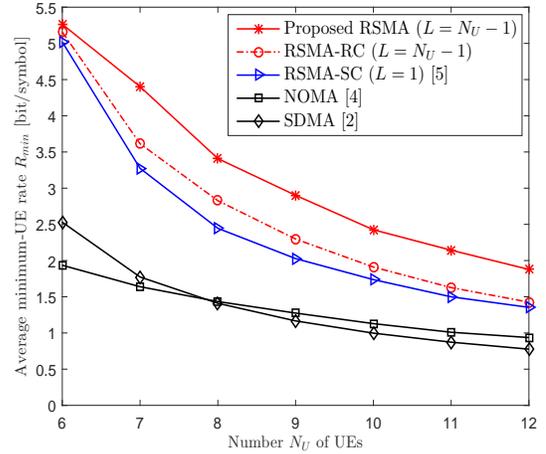}

\caption{{\footnotesize{}\label{fig:graph-asNU}Average minimum-UE rate $R_{\min}$
versus the number $N_{U}$ of UEs ($N_{R}=5$, $n_{R,i}=1$, $C_{i}=10$
and $P_{i}=43$ dBm).}}
\end{figure}

Fig. \ref{fig:graph-asNU} plots the average minimum-UE rate with
respect to the number $N_{U}$ of UEs for $N_{R}=5$, $n_{R,i}=1$,
$C_{i}=10$ and $P_{i}=43$ dBm. For small $N_{U}$, as mentioned
in Sec. \ref{sec:Introduction}, the performance of NOMA is worse
than that of SDMA, since the rates are limited to guarantee successful
SIC decoding. However, thanks to SIC, the NOMA scheme outperforms
SDMA as $N_{U}$ increases. Also, we observe that, while the performance
of the RSMA-RC scheme approaches that of RSMA-SC as $N_{U}$ increases,
the proposed RSMA scheme keeps a meaningful gain until $N_{U}=12$.
This comparison validates the importance of careful choice of the
sets $\mathcal{S}_{1},\mathcal{S}_{2},\ldots,\mathcal{S}_{L}$ for
common signals. It seems that the gain of the proposed RSMA scheme
slightly decreases for sufficiently large $N_{U}$. This is because
we focus on the minimum-UE rate which would be degraded for all the
compared schemes in overloaded systems with large $N_{U}$. The degradation
of the performance of all the schemes makes the gaps smaller.

\section{Conclusion\label{sec:Conclusion}}

We have studied the design of RSMA transmission for a C-RAN downlink
system. Specifically, we have proposed an efficient RSMA scheme that
uses a linearly increasing number of common signals. The decoding
UEs of each common signal are carefully chosen using hierarchical
clustering with inter-UE dissimilarity metric defined based on channel
directions. From numerical results, we observed that the proposed
RSMA scheme significantly outperforms the conventional SDMA and NOMA
schemes as well as conventional RSMA schemes that use a single common
signal or the same number of common signals with randomly chosen decoding
UEs.

\end{document}